\documentclass[aps,amsmath,amssymb,showpacs,twocolumn,floatfix,pra]{revtex4}

\usepackage{amssymb}
\usepackage{amsmath}
\usepackage{amscd}
\usepackage{graphicx}

\newcommand{\I}{\mathrm{I}}
\newcommand{\im}{\mbox{Im }}
\newcommand{\re}{\mbox{Re }}
\newcommand{\F}{{\mathcal{F}}}
\newcommand{\ab}{\hspace{-1.5mm}}
\newcommand{\V}{V_{\mathrm{dd}}}

\renewcommand{\vec}[1]{{\mathbf #1}}

\begin{document}


\title{Two-dimensional scattering and bound states of polar molecules in bilayers}

 \author{Michael Klawunn$^{1,2}$, Alexander Pikovski$^1$ and Luis Santos$^1$}
 \affiliation{
 \mbox{$^1$Institut f\"ur Theoretische Physik, Leibniz Universit\"at Hannover, Appelstr. 2, 30167 Hannover, Germany}\\
 \mbox{$^2$INO-CNR BEC Center and Dipartimento di Fisica, Universit\`a di Trento, 38123 Povo, Italy}}

\date{\today}

 \begin{abstract}
Low-energy two-dimensional scattering is particularly sensitive to the existence and 
the properties of weakly bound states. We show that interaction potentials $V(r)$ 
with vanishing zero-momentum Born approximation $\int dr r V(r)=0$ 
lead to an anomalously weak bound state which crucially modifies the two-dimensional 
scattering properties. This anomalous case is especially relevant in the context of polar molecules in bilayer arrangements.
 \end{abstract}

\maketitle
\section{Introduction}

Ultracold atomic gases are many-body systems, but most of their
fundamental properties originate from the underlying two-body problem, 
given by an interaction potential which is typically 
considered as short-range. The situation is completely different in 
dipolar gases due to the long-range character of the 
dipole--dipole interaction~\cite{Baranov2008,Lahaye2009}. 
Polar molecules with a potentially large electric dipole moment constitute a 
particularly exciting dipolar gas. However, achieving quantum 
degeneracy is handicapped by exothermic chemical reactions~\cite{Ospelkaus2010}. 
The latter may be avoided by confining the gas in two-dimensional geometries 
if the dipoles are polarized perpendicular to the trap plane, due to the 
repulsive character of the dipole--dipole interaction~\cite{Ni2010}. 

Bilayer (and in general multi-layer) arrangements 
of polar molecules offer the possibility of stability against 
inelastic reactions and give rise to inter-layer pairing 
due the dipole--dipole force~\cite{Klawunn2010,Pikovski2010,Demler}.
A dipole in layer $1$ interacts with a dipole in layer $2$, 
both dipoles oriented perpendicularly to the layers,
by the potential
\begin{equation}
\V(r)=U_0 \: (r^2-2)/(r^2+1)^{5/2}.
\label{eq:V}
\end{equation}
Here $r$ is the relative in-plane distance between the two dipoles,
the inter-layer distance $\lambda$ is set to $1$,
and $U_0$ is a positive dimensionless coupling constant.
We have $U_0=md^2/\hbar^2\lambda$ and the unit of energy is
$E_0=\hbar^2/m\lambda^2$, with $m$ the mass and $d$ dipole moment of the molecule.
This potential is attractive at short distances and repulsive at large distances,
fulfilling the peculiar condition $\int dr r \V(r)=0$, i.e. its 
zero-momentum Born approximation vanishes~\cite{Pikovski2010,Armstrong2010,Shih2009}.
Inter-layer interactions of a different type may be attained
in binary mixtures, where one of the species is confined in a bilayer
while the other moves freely.
The free species mediates a 2D interaction 
with a RKKY-like potential~\cite{Nishida2010}, which may as well
have a vanishing zero-momentum Born approximation.

Low energy 2D scattering, which determines the properties of 2D quantum gases~\cite{Petrov2003,Randeria1990,Miyake1983}, is particularly 
sensitive to the existence and properties of weakly bound states. 
Although 2D scattering~\cite{Bolle1984a,Bolle1984b,Gibson1986,Khuri2009,Chadan1998} and weakly bound states~\cite{Simon1976,Patil1980,Patil1982}
have been intensively studied, little is known for the case when $\int dr r V(r)=0$.
The binding energy for weakly coupled bound states in this case was calculated in 
Ref~\cite{Simon1976}. However, a detailed investigation of the binding energy at 
larger coupling and of the low-energy scattering properties is still lacking.

In this Brief Report we discuss the low-energy scattering and weakly bound states 
for radial potentials $V(r)$ satisfying $\int dr r V(r)=0$. 
We extend (using an alternative method) the expression derived
in Ref.~\cite{Simon1976} for the binding energy of the weakly bound state. 
We show that the presence of this anomalously weak bound state modifies significantly 
the scattering amplitude compared to the usual case of potentials with non-vanishing 
zero-momentum Born approximation. As an example, we specialize for the 
potential $\V$ appearing in bilayer gases of polar molecules and check the validity of the obtained analytical expressions using exact numerical calculations.

This paper is organized as follows.
In Sec.~\ref{sec:bound}, we introduce 
the Jost function formalism to
study the binding energy of weakly bound states
and evaluate the general expressions for $\V$.
In Sec.~\ref{sec:scattering} we discuss the modifications introduced in   
the two-dimensional scattering properties if $\int dr r V(r)=0$
and study this for $\V$ in detail.
Our conclusions are summarized in Sec.~\ref{sec:conclusions}.

\section{Weakly bound states}
\label{sec:bound}

Ref.~\cite{Simon1976} studied the bound states of the Schr\"odinger equation 
$\{-\nabla^2+V(\vec{r})\}\psi=\epsilon \psi$ for potentials of the form 
$V(\vec{r})=U_0 v(\vec{r})$, where as above $U_0$ denotes a positive dimensionless 
coupling constant characterizing the potential strength%
~\footnote{In Ref.~\cite{Simon1976} it was assumed that $\int d^2 r (1+|\vec{r}|^\delta)|v(\vec{r})|<\infty$, with $\delta>0$.}%
. For the case of weak coupling ($U_0 \to 0$) it was shown that a shallow 
bound state always exists if $\int d^2r v(\vec{r}) \leq 0$, 
but there is no bound state if $\int d^2r v(\vec{r}) > 0$.
Furthermore, it was shown 
that for $\int d^2r v(\vec{r}) < 0$ the binding energy 
of the shallow bound state is of the form
$\epsilon_b\sim -\exp[4\pi/\int d^2 r V(\vec{r})]$, as expected from e.g.~Ref.~\cite{LL3}. 
However, for our case of interest 
$\int d^2r v(\vec{r}) = 0$, the binding energy acquires the anomalous form 
$\epsilon_b\sim - \exp [1/cU_0^2]$, with  
\begin{equation}\label{cSimon}
c=\frac{1}{8\pi}\int d^2r \int d^2r' v(\vec{r}) \ln|\vec{r}-\vec{r}'|  v(\vec{r'}).
\end{equation}
It can be shown that $c<0$ for any $v(\vec{r})$.
For the potential $\V(r)$ appearing in a bilayer system of polar molecules, 
one obtains $c=-1/8$, and hence the binding energy becomes:
\begin{equation}\label{EbSimon}
\epsilon_b^{\mathrm{dd}} \sim -\exp\left\{-8/U_0^2\right\}.
\end{equation}
However, a numerical calculation of the binding energy (see below) shows that
this result is not very accurate even for very small $U_0$. 
This motivates us to derive a more accurate analytic expression 
for the binding energy which remains valid for larger $U_0$. 

The two-body scattering problem for a radially symmetric potential $V(r)$ in two dimensions
is described by the Schr\"odinger equation 
\begin{equation}\label{radialSE}
\left\{ - \left(\frac{d^2}{dr^2} + \frac{1}{r}\frac{d}{dr} \right) + V(r) \right\} \phi(r) = k^2 \phi(r),
\end{equation}
where all quantities are dimensionless and
$\phi(r)$ is the radial wavefunction. 
Only $s$-waves are considered since we are only interested in low-energy properties.
Following Ref.~\cite{Newton1986},
we employ the Jost function formalism to study the scattering problem and the shallow bound states. 
The definition of the Jost function $\F(k)$ in the 2D case may be found in Ref.~\cite{Newton1986}.
The properties of $\F(k)$ are similar
to those of the Jost function for the 3D case~\cite{Newtonbook}. 

The scattering phase shift $\delta(k)$ is related to $\F(k)$ by
\begin{equation}
\tan\delta(k)=-\frac{\im \F(k)}{\re \F(k)},
\label{eq:delta} 
\end{equation}
the scattering amplitude is 
\begin{equation}
f(k)=\frac{\tan\delta(k)}{1-i\tan\delta(k)},
\end{equation}
and 
$\sigma = (4/k) |f(k)|^2=(4/k) \sin^2 \delta(k)$
is the total 2D s-wave cross-section.
For complex $k$, the zeros of the Jost function on the positive 
imaginary axis, $\F(i\alpha)=0$ with $\alpha>0$, are the bound states of the 
potential with binding energy $\epsilon_b=-\alpha^2$.

The following integral representation of the Jost function $\F(k)$ will be employed below:
\begin{equation}\label{Jostfunc}
\F(k)=1+e^{i\pi/4}\sqrt{\frac{\pi}{2k}}
\int_0^\infty dr \sqrt{r}\, V(r) J_0(kr)f_0(kr),
\end{equation}
where $f_0(k,r)$ satisfies the integral equation

\begin{equation}\label{Jostsol}
f_0(k,r)\!=\!\sqrt{\frac{i \pi kr}{2}}H_0(kr)
+\!\int_r^\infty \ab\ab\! ds\, g(k,r,s) V(s) f_0(k,s),\!
\end{equation}
with
\begin{equation}\nonumber
g(k,r,r')=\frac{\pi}{2} \sqrt{rr'}\left[J_0(kr)Y_0(kr')-J_0(kr')Y_0(kr) \right].
\end{equation}
In the previous expressions $J_0$, $Y_0$ are Bessel functions, and 
$H_0$ is the Hankel function of the first kind.

\begin{figure}
\begin{center}
\includegraphics[width=0.3\textwidth,angle=270]{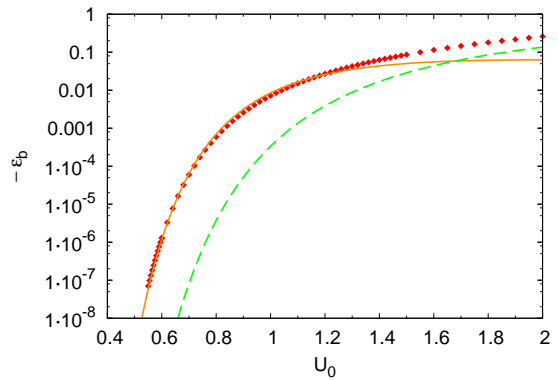}
\caption{(Color online) Binding energy for $\V$ as a function of the dipole strength $U_0$
calculated numerically (dots), from equations~(\ref{Eb}) (solid) 
and~\eqref{EbSimon} (dashed).}
\label{fig:1}
\end{center}
\vspace*{-0.7cm}
\end{figure}

In the following we are interested in determining the weakly bound states. To this aim 
we expand the Jost function for small $k$:
\begin{equation}\label{Jostlog}
\F(k)=A \ln k + B-i\frac{\pi}{2}A,
\end{equation}
where $A$, $B$ are real constants independent of $k$.
These constants can be represented by infinite series of the form 
$A=\sum A_j$, $B=1-\sum B_j$
\footnote{We always assume $A\neq 0$. The exceptional case $A=0$ (cf.~\cite{Newton1986})
is the two-dimensional analogue of zero-energy resonance in three dimensions.}
. The first terms of the $A_j$ series are of the form:
\begin{eqnarray}
A_1\!\!&=&\!\!-\int_0^\infty \ab dr r V(r)\\
A_2\!\!&=&\!\!-\int_0^\infty \ab dr r V(r)
\int_{r}^\infty \ab ds s V(s) \ln\left(\frac{s}{r}\right),
\end{eqnarray}
whereas those of the $B_j$ series are:
\begin{eqnarray}
B_1\!\!&=&\!\!
\int_0^\infty \ab dr r V(r)\ln(r)
-C_1
\\
B_2\!\!&=&\!\! 
\int_0^\infty \ab\ab dr r V(r)
\int_{r}^\infty \ab\ab ds s V(s) \ln\left(\frac{s}{r}\right)\ln(s)
-C_2,
\end{eqnarray}
with $C_i=A_i \ln (e^\gamma/2)$ and $\gamma\approx 0.577$ the Euler constant.

As mentioned above, 
the bound states are given by the zeros of $\F(k)$ on the positive imaginary axis.
Using (\ref{Jostlog}) with $k=i\alpha$ we hence obtain 
the expression of the binding energy:
\begin{equation}\label{Eb_general}
\epsilon_b=-\exp\left\{-2\frac{B}{A}\right\}.
\end{equation}
Note that Eq.~\eqref{Eb_general} is valid as long as the 
binding energy is small enough such that the logarithmic
term dominates the Jost function.
For potentials with $\int dr r V(r)<0$, 
it is sufficient to take $A \approx A_1$ and $ B \approx 1$, recovering the 
expression $\epsilon_b\sim \exp(4\pi/\int d^2r V(r))$ for small $U_0$.

However, for the case of potentials such that 
$\int dr r V(r)=0$, we have $A_1=0$, 
and the first non-vanishing term is $A \approx A_2$, $B \approx 1$, providing 
an alternative derivation of Eq.~\eqref{EbSimon}.
A more precise formula is obtained by including higher-order terms:
\begin{equation}\label{Eb_general_2}
\epsilon_b=-\exp\left\{-2\:\frac{1-B_1-B_2 -\ldots}{A_1+A_2 +\ldots} \right\} .
\end{equation}

For the case of the inter-layer dipole--dipole potential $\V$, 
the integrals can be carried out analytically to find a corrected 
expression for the binding energy
\begin{equation}\label{Eb}
\epsilon_b^{\mathrm{dd}}\!\simeq\! -\exp\left\{-\frac{8}{U_0^2}\left[
    1\!-\!U_0\!+\!\frac{U_0^2}{4}\left(\frac{5}{2}\!+\!\ln \tfrac{e^\gamma}{2} \right)
 \right]  \right\} .
\end{equation}

Figure~\ref{fig:1} compares the numerical result 
for the binding energy for $\V$ (obtained directly from the 2D Schr\"odinger equation) 
with the analytical expressions of Eq.~\eqref{EbSimon} and \eqref{Eb}. Note that whereas 
Eq.~\eqref{EbSimon} provides a relative inaccurate approximation even at rather low $U_0$, 
the newly derived expression~\eqref{Eb} is in excellent agreement with the numerics, 
all the way to $U_0 \lesssim 1.2$.

Finally, we note that for large $U_0$ the binding energy for $\V$ can be determined by a variational 
calculation, giving~\cite{Yudson1997} $\epsilon^{\mathrm{dd}}_b\approx-2 U_0 + 4\sqrt{3U_0/2} - 15/4$ ,
which coincides with the numerics only for $U_0 \gtrsim 5$.

\section{Scattering phase shift}
\label{sec:scattering}

\begin{figure}
\vspace*{-0.2cm}
\begin{center}
\includegraphics[width=0.3\textwidth,angle=270]{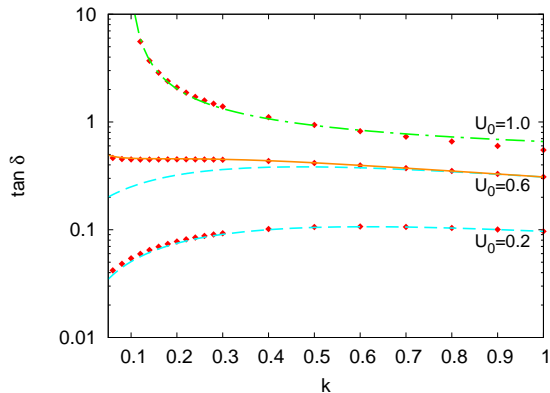}
\vspace*{-0.2cm}
\caption{(Color online) Scattering phase shift for $\V$ as a function of $k$ for 
different $U_0$ calculated numerically (dots) compared to 
the log-behavior (\ref{ampl-log}) for $U_0=1.0$ (dash-dotted, green),
to formula (\ref{amplJost}) for $U_0=0.6$ (solid, orange)
and to the second Born approximation
(\ref{ampl-born2}) for $U_0=0.6$ and $U_0=0.2$ (dashed, cyan).
}
\label{fig:2}
\end{center}
\vspace*{-0.7cm}
\end{figure}

We have shown above that the Jost function formalism is particularly useful for the analysis of weakly bound states. 
In this section we employ this formalism for the study of 2D scattering
and in particular for the calculation of the $s$-wave 
scattering phase shift $\delta(k)$.

An approximate expression of the Jost function $\F(k)$ for small $U_0$ is obtained
by iterating twice the integral equation~(\ref{Jostsol}). 
Note that we keep all orders in $k$. The resulting scattering phase shift 
follows from the relation~(\ref{eq:delta}):
\begin{equation}\label{amplJost}
\tan\delta(k)
\!=\!\!\frac{\!-\tfrac{\pi}{2} \I_{JJ}(k)\!-\!\tfrac{\pi^2}{4}\left( \I_{JJ,JY}(k)
\!-\!\I_{JY,JJ}(k)\right)}{\!1\!-\!\tfrac{\pi}{2}\I_{JY}(k)
\!-\!\tfrac{\pi^2}{4}\left(\I_{JJ,YY}(k)
\!-\!\frac{1}{2}\I_{JY}^2(k)\right)},
\end{equation}
where we have introduced the notation
\begin{eqnarray}
&&\!\!\!\!\!\!\!\!\!\!\!\!\!\!I_{FG}\!=\!\!\!\int_0^\infty\!\!\!\! dr r V(r) F(r) G(r), \\
&&\!\!\!\!\!\!\!\!\!\!\!\!\!\!I_{FG,PQ}\!\!=\!\!\!\int_0^\infty\!\!\!\! dr r V(r) F(r)G(r)
\!\! \int_r^\infty \!\!\!\! ds s V(s)\! P(s) Q(s), 
\end{eqnarray}
and $J$, $Y$ stand for $J_0(kr)$ and $Y_0(kr)$.


For small $k$ it is possible to simplify Eq.~(\ref{amplJost}).
Employing the logarithmic expression~\eqref{Jostlog}, 
the relation~\eqref{eq:delta}, and the expression for the 
binding energy~(\ref{Eb_general}), we recover the well known
logarithmic expression (see e.g. \cite{Randeria1990}) 
\begin{equation}\label{ampl-log}
\tan \delta(k)= \left[ \frac{1}{\pi}\ln\frac{k^2}{|\epsilon_b|} \right]^{-1},
\end{equation}
characteristic of 2D scattering, which relates the scattering 
shift and the binding energy of the weakly bound state. 
However, for the case $\int dr r V(r)=0$ the binding energy  
$|\epsilon_b|$ can become anomalously small, and hence 
the expression at the right-hand side
of Eq.~(\ref{ampl-log}) can become very small for reasonable $k$.
In this case, it is not any more the leading term for the low-energy scattering.

\begin{figure}
\begin{center}
\includegraphics[width=0.38\textwidth]{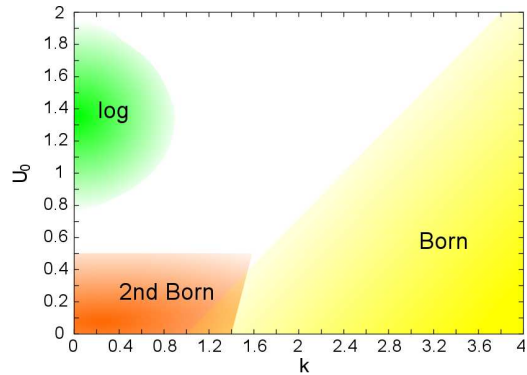}
\vspace*{-0.2cm}
\caption{(Color online) Qualitative sketch of the regimes of $k$ and $U_0$
where the scattering phase shift for $\V$ can be approximated by the 
logarithm Eq.~(\ref{ampl-log}) (green), the Born approximation Eq.~(\ref{ampl-born}) (yellow), 
and the second Born approximation Eq.~(\ref{ampl-born2}) (orange).
}\label{fig:3}
\end{center}
\vspace*{-0.7cm}
\end{figure}

On the other hand, for large enough $k$ and small $U_0$,
the first integral $\I_{JJ}(k) $ in Eq.~(\ref{amplJost}) dominates, and we recover, as expected for sufficiently large $k$, the Born approximation.
Therefore, formula~(\ref{amplJost}) interpolates smoothly between the correct low-energy and the correct
high-energy behavior.
This suggests that it may be valid, at least qualitatively, even for large $U_0$.
By expanding Eq.~(\ref{amplJost}) in powers of $U_0$, one recovers the Born series,
which reads to second order:
\begin{equation}\label{ampl-born2}
\tan\delta(k)=- \frac{\pi}{2} I_{JJ} - \frac{\pi^2}{2} \, I_{JJ,JY}.
\end{equation}

It is interesting to discuss these results for the case of the 
interlayer dipole--dipole potential $\V$. 
We have computed the scattering phase-shift numerically from the Schr\"odinger equation 
and compared it to the results of Eq.~(\ref{amplJost}), 
obtaining that Eq.~(\ref{amplJost}) provides the correct scattering phase-shift with excellent 
accuracy, at least in the range $0.03 \le k \le 5$ and $0.05 \le U_0 \le 2.0$.
Hence Eq.~\eqref{amplJost} is a good approximation not only for $U_0 \ll 1$, but also for $U_0 \sim 1$. 
Of course, if the interactions are too large ($U_0 \gg 1$), the contributions from higher iterations 
of the integral equation~(\ref{Jostsol}) become more important and Eq.~(\ref{amplJost}) looses its accuracy.

Figure~\ref{fig:2} compares the numerical results for the scattering phase-shift for $\V$
with the limiting cases provided by expressions~(\ref{ampl-log}) and ~(\ref{ampl-born2}) for small $k$ and different values of $U_0$.
It is seen that for $U_0 \sim 1$, the scattering phase-shift is best approximated by the logarithmic
expression~(\ref{ampl-log}), and for $U_0 \ll 1$ by the second Born approximation~(\ref{ampl-born2}). 
For intermediate values of $U_0$ none of the limiting cases is accurate and the full expression~(\ref{amplJost}) must be used.
We sketch in Fig.~\ref{fig:3} qualitatively the regimes of $k$ and $U_0$, where the logarithm~(\ref{ampl-log}),
the first Born approximation~(\ref{ampl-born}), and the second Born approximation~(\ref{ampl-born2}) 
are good approximations, as obtained by comparison with the numerical solution.
Note that, excluding unreasonably small $k$,
the logarithmic form~(\ref{ampl-log}) is just valid for $k<1$ and 
the window $0.7\lesssim U_0 \lesssim 2.0$.

Finally, we note that the first Born approximation 
for $\V$ can be evaluated exactly analytically:
\begin{equation}\label{ampl-born}
\tan\delta^{\mathrm{dd}}(k)\!\simeq\!-\frac{\pi}{2} U_0 \left[-\frac{4k}{\pi}\!-\!2 k \left(\mathbf{L}_1(2k)-I_1(2k) \right) \right],
\end{equation}
where $\mathbf{L}_1$ is the modified Struve function.
The second Born approximation (\ref{ampl-born2}) can be expanded for small $k$
\begin{equation}
\tan\delta^{\mathrm{dd}}(k)\approx 2 U_0 k -\pi U_0 k^2 +\tfrac{1}{8} U_0^2\pm\ldots ,
\end{equation}
which gives a maximum in the scattering phase shift
at $k\approx1/\pi$ as observed in the numerical results.

\section{Conclusions}
\label{sec:conclusions}

In conclusion, two-dimensional radial interaction potentials $V(r)$ with a vanishing 
zero-momentum Born approximation, $\int dr r V(r)=0$, result in interesting physics
crucially different from purely attractive or purely repulsive potentials.
Using the Jost function formalism, we have derived an expression for 
the binding energy as a function of the potential strength $U_0$, 
which remains accurate for a wide regime of $U_0$ values.
Moreover, we have investigated the scattering amplitude in different parameter regimes.
In particular, we have shown a significant deviation of the scattering
behavior in comparison with potentials with $\int dr r V(r)\neq 0$, due to the 
anomalously low binding energy of the weakly bound state. 

These results are of particular importance in the physics of two-dimensional systems, 
and more specifically on two-dimensional ultracold gases. 
Standard theories, in particular the theory of BCS-BEC crossover~\cite{Randeria1990,Miyake1983},
are based on the fact that the scattering amplitude
possesses the logarithmic dependence~(\ref{ampl-log}). 
These results are therefore modified if the potential has a vanishing zero-momentum Born approximation. 
This has particularly important consequences for the properties of a gas of
polar Fermi molecules confined in a bilayer geometry, including 
inter-layer pairing~\cite{Pikovski2010}.

\acknowledgments
We thank A.~Recati and G.V.~Shlyapnikov for helpful discussions.
We acknowledge the support of the DFG (QUEST Cluster of Excellence).


\begin{thebibliography}{99}


\bibitem{Baranov2008} M. A. Baranov, Phys. Rep. {\bf 464}, 71 (2008).

\bibitem{Lahaye2009} T. Lahaye, C. Menotti, L. Santos, M. Lewenstein and T. Pfau, Rep. Prog. Phys. {\bf 72}, 126401 (2009).

\bibitem{Ospelkaus2010}
S. Ospelkaus {\it et al.}, Science {\bf 327}, 853 (2010).

\bibitem{Ni2010}
K.-K. Ni {\it et al.}, Nature {\bf 464}, 1324 (2010).

\bibitem{Klawunn2010} M. Klawunn, J. Duhme, and L. Santos,
Phys. Rev. A {\bf 81}, 013604 (2010). 

\bibitem{Demler}
A. C. Potter, E. Berg, D.-W. Wang, B. I. Halperin and E. Demler,
{\tt arXiv:1007.5061}.

\bibitem{Pikovski2010}
A. Pikovski, M. Klawunn, G. V. Shlyapnikov and L. Santos,
{\tt arXiv:1008.3264}.

\bibitem{Shih2009}
S.-M.~Shih and D.-W.~Wang,
Phys. Rev. A {\bf 79}, 065603 (2009). 

\bibitem{Armstrong2010}
J. R. Armstrong {\it et al.}, 
EPL {\bf 91}, 16001 (2010).

\bibitem{Nishida2010}
Y. Nishida,
Phys. Rev. A {\bf 82}, 011605 (2010).

\bibitem{Petrov2003}
D. S. Petrov, M. A. Baranov and G. V. Shlyapnikov,
Phys.~Rev.~A {\bf 67}, 031601(R) (2003).

\bibitem{Randeria1990}
M. Randeria, J.-M. Duan and L.-Y. Shieh,
Phys.~Rev.~B {\bf 41}, 327 (1990);

\bibitem{Miyake1983} K. Miyake, Prog. Theo. Phys. {\bf 69}, 1794 (1983).

\bibitem{Bolle1984a}
D. Boll\'e and F. Gesztesy,
Phys. Rev. Lett. {\bf 52}, 1469 (1984).

\bibitem{Bolle1984b}
D. Boll\'e and F. Gesztesy,
Phys. Rev. A {\bf 30}, 1279 (1984).

\bibitem{Gibson1986}
W.G. Gibson,
Phys. Lett. A {\bf 117}, 107 (1986).

\bibitem{Khuri2009}
N. N. Khuri, A. Martin, J.-M. Richard and T. T. Wu,
J. Math. Phys. {\bf 50}, 072105 (2009).

\bibitem{Chadan1998}
K. Chadan, N. N. Khuri, A. Martin and T. T. Wu,
Phys. Rev. D {\bf 58}, 025014 (1998).

\bibitem{Simon1976}
B. Simon, Annals of Physics {\bf 97}, 279 (1976).

\bibitem{Patil1980} S. H. Patil, Phys. Rev. A {\bf 22}, 2400 (1980).
\bibitem{Patil1982} S. H. Patil, Phys. Rev. A {\bf 25}, 2467 (1982).

\bibitem{LL3}
L. D. Landau and E. M. Lifshitz, {\em Quantum Mechanics},
Butterworth-Heinemann, Oxford, 1977. 

\bibitem{Newton1986}
R. G. Newton,
J. Math. Phys. {\bf 27}, 2720 (1986).

\bibitem{Newtonbook}
R. G. Newton,
{\em Scattering theory of waves and particles}, New York, McGraw-Hill, 1966.

\bibitem{Yudson1997}
V. I. Yudson, M. G. Rozman, and P. Reineker,
Phys. Rev. B {\bf 55}, 5214 (1997).

\end{thebibliography}
\end{document}